%% file: main.tex
\documentclass{Interspeech2024}

\interspeechcameraready

\title{ASTRA: Aligning Speech and Text Representations for Asr without Sampling}

\name{Neeraj Gaur*, Rohan Agrawal*, Gary Wang, Parisa Haghani, Andrew Rosenberg, Bhuvana Ramabhadran}
\address{Google, U.S.A}
\email{}

\newcommand\blfootnote[1]{%
  \begingroup
  \renewcommand\thefootnote{}\footnote{#1}%
  \addtocounter{footnote}{-1}%
  \endgroup
}

\begin{document}

\maketitle

\blfootnote{\textsuperscript{*} Equal contribution}
\blfootnote{To appear in Interspeech 2024}

\begin{abstract}
    \input{inputfiles/abstract}
\end{abstract}

\section{Introduction}
    \input{inputfiles/introduction}
    
\section{Background}
    \input{inputfiles/background}
    
\section{ASTRA model architecture}
    \input{inputfiles/sampling_free}

\section{Experimental setup}
    \input{inputfiles/experiments}
    
\section{Results}
    \input{inputfiles/results}
    
\section{Conclusion}
    \input{inputfiles/conclusion}    

\bibliographystyle{IEEEtran}
\bibliography{mybib}

\end{document}

%% file: inputfiles/abstract.tex
This paper introduces ASTRA, a novel method for improving Automatic Speech Recognition (ASR) through text injection. Unlike prevailing techniques, ASTRA eliminates the need for sampling to match sequence lengths between speech and text modalities. Instead, it leverages the inherent alignments learned within CTC/RNNT models. This approach offers the following two advantages, namely, avoiding potential misalignment between speech and text features that could arise from upsampling and eliminating the need for models to accurately predict duration of sub-word tokens. This novel formulation of modality (length) matching as a weighted RNNT objective matches the performance of the state-of-the-art duration-based methods on the FLEURS benchmark, while opening up other avenues of research in speech processing.

%% file: inputfiles/introduction.tex
Text-only data can be used to boost the performance of ASR models \cite{bapna2021slam, chen2022maestro}. Moreover, large multi-modal models \cite{team2023gemini, achiam2023gpt} have ushered in an era of exciting new advancements in various domains. Audio-text multi-modality, in particular, has shown great promise in improving the quality of Automatic Speech Recognition (ASR) systems \cite{le2023pre, tsunoo2023decoder} in various settings like low resource languages \cite{chen2023maestrou, bapna2022mslam, 10095218}, spoken language understanding \cite{bapna2021slam, thomas2022towards}, recognition of named entities and alphanumerics \cite{blau2023textinject}, among others.

Modality matching techniques, where a consistency loss is enforced between speech and text representations \cite{chen2022maestro,wang2023understanding,saeki2023virtuoso}, are often used to boost performance of such speech-text multi-modal systems. Coupled with pre-trained foundation models, modality matching can unlock a number of exciting zero-shot/few-shot capabilities \cite{wang2023slm}. However, these techniques typically require up-sampling the text sequence, either by using a fixed duration or a learned duration model, to roughly match the lengths of the audio sequences, introducing potential complexities. When up-sampling the text sequence, and using the up-sampled sequence for modality matching, one runs the risk of aligning text tokens with the speech sequence corresponding to silences or noise, or even to parts of the speech sequence corresponding to emission of other text tokens. Moreover, the quality of the duration model can have a big impact on the overall performance of the system \cite{sainath2022joist}, and it has been observed that the duration model based approaches are very dependent on the domain of the duration model matching the test domain \cite{wang2023understanding}.

This paper introduces ASTRA, a novel method that addresses the limitations of sampling-based text injection approaches for RNNT based models. Our contribution centers on two key innovations:
\begin{itemize}
    \item Novel formulation of modality matching: Eliminating the need for explicit length matching, ASTRA leverages the implicit alignments learned by RNNT models.

    \item Novel approximation: We show that an additive loss over an alignment path can be viewed as a weighted RNNT loss opening the way for novel applications.
\end{itemize}

While we present our discussion in terms of RNNT for ease of exposition, the formulation holds for CTC models as well.

%% file: inputfiles/background.tex
In recent years e2e models have become dominant in many fields. A popular e2e model for ASR is RNNT \cite{graves2012sequence}. E2E models like RNNT allow greater flexibility in gathering training data since they don't need explicit alignments. This lack of  alignments poses a challenge for text injection methods when it comes to enforcing consistency between speech and text embeddings since the sequence lengths are mismatched. Prior work \cite{chen2022maestro, saeki2023virtuoso} has got around this issue by generating alignments between text and speech on the fly, learning a duration model by first explicitly aligning speech with text using  viterbi alignment, and up-sampling the text sequence based on the learned duration model to match the length of the speech sequence. In~\cite{sainath2022joist}, the authors compared the approaches presented in~\cite{bapna2021slam, chen2022maestro} exploring fixed length, random length up-sampling and suggested learning the distribution of each sub-word unit as alternate to enforce durations. Some other works have also tried to get around the issue of a lack of alignment by first explicitly aligning speech and text use dynamic time warping \cite{peyser2023improving} or optimal transport \cite{le2023pre}.

While a comprehensive overview of an RNNT model is beyond the scope of this paper, it is sufficient to note that the RNNT model solves the problem of efficiently computing the full posterior i.e.,
\[ p(Y/X) = \sum_{a \in B(Y)} p(a)\]
where, $Y$ is the target sequence, text in our case, and $X$ is the input sequence, audio in our case. $B(Y)$, is the set of all possible valid alignments admitted by the RNNT lattice. For a full description of RNNT the reader is referred to \cite{graves2012sequence}.

%% file: inputfiles/sampling_free.tex
\subsection{Model architecture and losses}
    \begin{figure}[h]
      \centering
      \includegraphics[width=\linewidth]{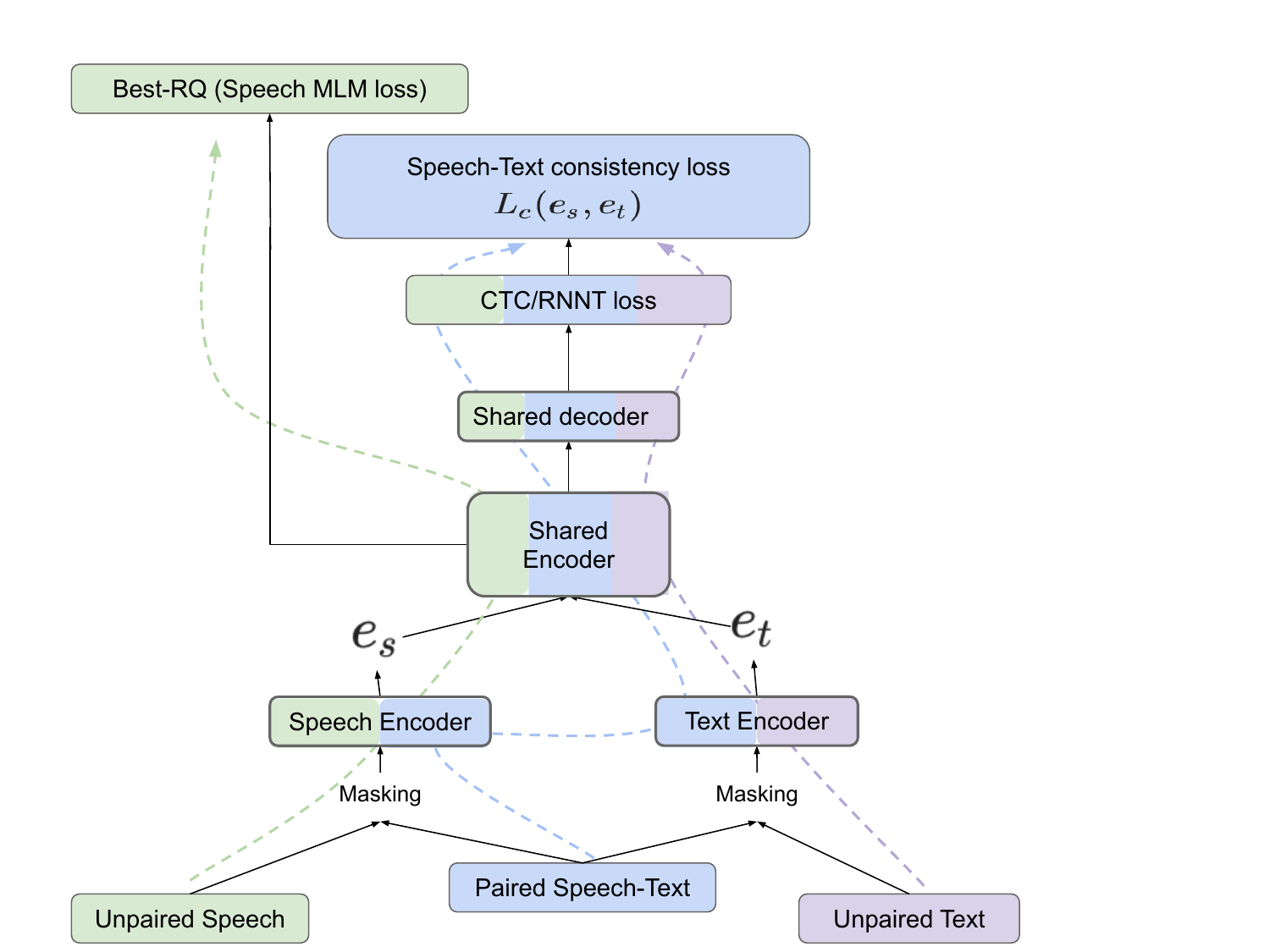}
      \caption{Model architecture with losses. Light green represents all parts of the model and that are active for unpaired speech data. Similarly, light blue and purple show parts active for paired speech-text data and unpaired text data respectively.}
      \label{fig:model_arch}
    \end{figure}
    \vspace{2.5mm}
    The complete model architecture is shown in Figure~\ref{fig:model_arch}. Our architecture is similar to the architecture used in \cite{chen2022maestro}. As in \cite{chen2022maestro}, the model consists of a speech encoder, a text encoder, a shared encoder and a shared decoder. 
    
    With unpaired speech, we mask the input and pass it through the speech and shared encoders and we optimize with BEST-RQ \cite{chiu2022selfsupervised}, which is a self-suerpvised masked language model objective used with quantized speech features where the quantization is done by randomly projecting to discrete bins.
    
    When learning with unpaired text, we extract embeddings based on the input text and pass it through the text and shared encoders, through the shared auto-regressive decoder and optimize the RNNT objective i.e. we maximize $p(t/e_t)$, probability of the text $t$, given the (optionally masked) text features $e_t$, similar to Eqn 2 in \cite{chen2022maestro}. Note that unlike \cite{chen2022maestro}, we do not learn a duration model, nor do we up-sample the text embeddings.
    
    With paired data, the input audio features are passed through the speech and shared encoders and the shared decoder and the ASR loss (RNNT loss) is computed. In addition to the ASR loss, we also add a modality-matching loss to enforce consistency between the speech and text embeddings. The text embedding sequence is obtained by passing the unmasked reference text through the text encoder. For the speech embeddings sequence we use the output of the speech encoder. Unlike \cite{chen2022maestro} the consistency loss is computed between the speech and text embedding sequences without matching their lengths. We instead use the alignments learned by the RNNT model itself to enforce consistency. Details of the consistency loss are presented next.
    
\subsection{Learned alignment speech-text consistency}
\label{ssec:learned-alignment}
    Modality matching through a consistency loss has shown to improve the quality of text injection\cite{chen2022maestro, wang2023understanding, peyser2023improving, saeki2023virtuoso}. Consistency aims to bring speech embeddings ($e_s$), the output of a speech encoder, and text embeddings ($e_t$), the output of the text encoder, closer.
    
\begin{figure}[t]
      \centering
      \includegraphics[width=\linewidth]{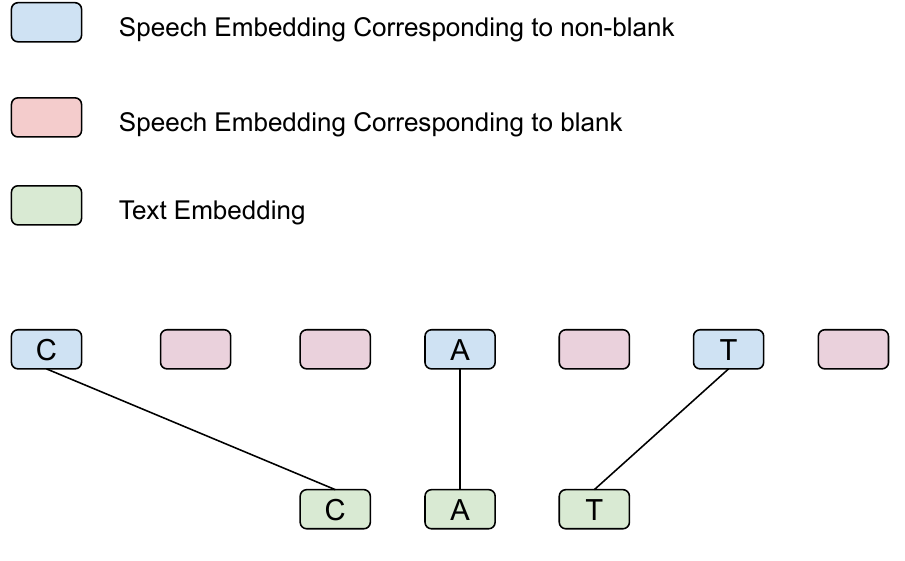}
      \caption{Toy example of a speech alignment which consists of non-blank frames, shown in blue, and blank frames, shown in red. As shown consistency is only enforced between speech frames corresponding to non-blank tokens and corresponding text embedding}
      \label{fig:align_speech_text}
    \end{figure}

    As mentioned earlier, one issue in modality matching is the length mismatch between speech and text modalities. Notice however, that if we were given the alignment between speech and text, enforcing consistency would be straightforward. As shown in Figure~\ref{fig:align_speech_text}, we can simply ignore the speech frames corresponding to an emission of a blank symbol and only apply consistency loss between the speech embeddings corresponding to the non-blanks and the corresponding text embeddings. This insight, that for a particular alignment it is enough to enforce consistency on only the non-blank frames, allows us to extend the consistency loss formulation when we do not have an explicit alignment. We will, instead, make use of the implicit alignment learned by an RNNT itself to enforce consistency between speech and text embeddings. Next, we show that the explicit alignment based upsampling can be compactly replaced by the marginalization step in the proposed approach.

    \vspace{2.5mm} 

    For a particular alignment $a$, the consistency loss can be computed as:
    \begin{equation}
        L_{ca} = \sum_{(k,u)\in a} L (e_s (k,u), e_t (u) )
     \end{equation} 
     where, $L(.,.)$ is a pointwise consistency loss (we use Mean Absolute Error as our consistency loss), $e_s (k,u)$ is the output of the speech encoder at frame $k$ corresponding to emission of non-blank symbol $u$, and $e_t (u)$ is the output of the text encoder corresponding to the $u$-th symbol in the text sequence. 
     \newline
     As noted above, we let
     \begin{equation} 
        L (e_s (.,\epsilon), e_t (.) ) =  0, 
     \end{equation} 
     i.e. we only enforce consistency loss on the speech frames corresponding to non-blank emissions.
    
    Summing up over all alignments gives the overall consistency loss as follows:
    \begin{equation}
        L_c = \frac{1}{p(Y/X)} \sum_{a \in B(Y)} p(a) * L_{ca}
    \end{equation} 
    i.e., \begin{equation}
        L_c = \frac{1}{p(Y/X)} E_{a \in B(Y)}[ L_{ca} ]
    \end{equation}
    where, $X$ is the speech sequence, $Y$ is the text sequence, $p(Y/X)$ is the (full-sum) probability assigned to $Y$ given $X$, $B(Y)$ is the set of all possible valid RNNT alignments and $p(a)$ is the probability assigned to a particular alignment by the RNNT.
    
    While, $L_c$ can be efficiently computed using the expectation semi-ring as shown by \cite{eisner2002parameter}. We will instead optimize $\widehat{L_c}$, which is defined as follows: 
    \begin{equation}
        \widehat{L_c} := \frac{1}{p(Y/X)} log(E_{a \in B(Y)} [e^{L_{ca}}] )
    \end{equation}
    The normalizing term $\frac{1}{p(Y/X)}$ is the full likelihood of the sequence given by the RNNT. This allows us to view consistency loss as a weighted RNNT loss.
    \vspace{2.5mm}
    \newline
    Using the definition of $L_{ca}$ and expanding the second term we get:
    \vspace{2.5mm}
    \begin{equation}
        E_{a \in B(Y)} [e^{L_{ca}}]= \sum_{a \in B(Y)} \prod_{(k,u) \in a}p(k,u) * e^{L(e_s(k,u), e_t(u))}
    \vspace{2.5mm}
    \end{equation} 

\begin{figure}[th]
      \centering
      \includegraphics[width=\linewidth]{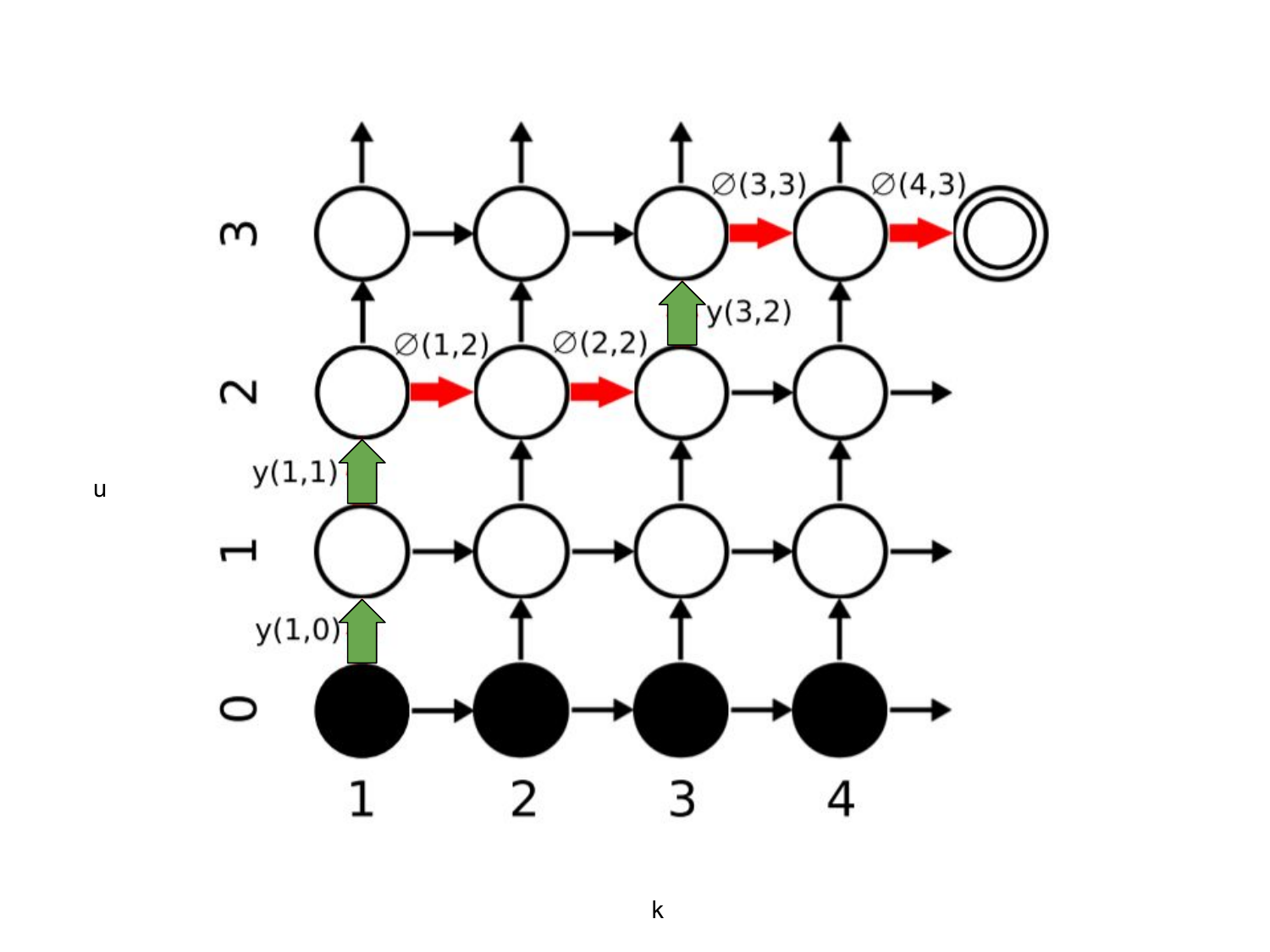}
      \caption{RNNT lattice \cite{graves2012sequence} where vertical transitions represent non-blank emissions and horizontal transitions represent blank emissions. Also shown is the path for one alignment through the lattice. Here green arrows represent weighted non-blank transitions and red arrows represent weighted blank transitions.}
      \label{fig:weighted_rnnt}
    \end{figure}

    This can be viewed as a weighted RNNT loss  where the non-blank transitions, represented by the green arrows in Figure~\ref{fig:weighted_rnnt}, are weighted by the corresponding pointwise consistency loss term.
    
    We will now see that minimizing $\widehat{L_c}$ is equivalent to minimizing an upper bound on $L_c$.
    \newline\vspace{2.5mm}
    From Jensen's inequality we get:
    \begin{equation}
        E_{a \in B(Y)} [e^{L_{ca}}] \geq e^{E_{a \in B(Y)} [L_{ca}] }
    \end{equation} 
    Therefore, \begin{equation}
        log(E_{a \in B(Y^*)} [e^{L_{ca}}]) \geq E_{a \in B(Y)} [L_{ca}] 
    \end{equation}
    \[ \Rightarrow \widehat{L_c} \geq L_c \]
    \vspace{2mm}
    
    Hence, we see that any loss which can be expressed as additive over a particular alignment can be viewed as a weighted RNNT loss and efficiently applied over all alignments. This allows us to treat blank and non-blank transitions differently while still being able to optimize the resulting loss efficiently. We believe this paves the way for other novel applications such as integrating with LMs or biasing, where one may want to treat certain transitions differently from others.

%% file: inputfiles/experiments.tex
\renewcommand{\arraystretch}{1.5} 
\begin{table}[!htp]\centering
    \caption{Description of datasets used}\label{tab:ablation}
    \scriptsize
    \begin{tabular}{p{1.5cm} p{1.5cm} p{1.5cm} p{1.5cm}}\hline
    \textbf{Dataset} &\textbf{Modality} &\textbf{Data Size} &\textbf{\# of Languages} \\\hline \hline
    YT-56-U &Audio &One million hours &56\\ \hline
    mC4 &Text &6.3T tokens &101 \\ \hline
    Fleurs train &Speech-Text &987 hours &102 \\ \hline
    Fleurs dev &Speech-Text &120 hours &102 \\ \hline
    Fleurs test &Speech-Text &283 hours &102 \\ \hline
    \end{tabular}
    \vspace{2.5mm}
\end{table}

\subsection{Architecture}
    The ASR network is an RNNT network \cite{graves2012sequence} consisting of a 2 layer LSTM decoder and a stack of 24 Conformer encoder blocks \cite{gulati2020conformer}. Each of these conformer blocks contains multi-headed self attention \cite{vaswani2023attention}, depth-wise convolution and feed-forward layers. All models are trained on 80-dimensional log-mel filter bank coefficients. Our experiments use 4096 sentence-piece targets. Our model has approximately 300M parameters. All ASR models are trained on Google TPU V3 cores \cite{jouppi2020domain}. We use Adam optimization and cap the norm of the gradient to 5. We use a transformer learning rate schedule \cite{vaswani2023attention}. We split the 24 conformer encoder blocks into speech encoder and shared encoder as can be seen in Figure \ref{fig:model_arch}. Similar to \cite{chen2022maestro}, the speech encoder contains 6 conformer blocks and the shared encoder has the remaining 18 conformer blocks. The Text encoder consists of a text embedding layer followed by 4 stacked Conformer blocks. 
\subsection{Pretraining}
    We pretrain our models on YT-56-U, which consists of one million hours of audio from "speech-heavy" Youtube videos. The data is segmented by a Voice Activity Detection model and non speech segments are removed. More details of this dataset are in \cite{zhao2024usmscd}. We use BEST-RQ \cite{chiu2022selfsupervised} for self-supervised BERT-style pretraining to learn from speech only data. The pretraining step uses the same encoder architecture as described in the previous sub-section.
\subsection{Training and evaluation}
    We continue training the pretrained encoder on supervised speech. Once speech embeddings stabalize, we enable the alignment loss, and after a few steps, we enable text only loss. We use the FLEURS dataset \cite{conneau2022fleurs} for ASR training and evaluation. This dataset contains around 12 hours of supervised speech data for 102 languages. For unspoken text, we use mC4 which is drawn from the public Common Crawl web scrape and spans 101 languages. More details about mc4 are in \cite{xue2021mt5}

    For evaluation we report the average CER over all 102 locales on the FLEURS test set. We decode using greedy decode strategy.
\subsection{Baselines}
    {\bf Vanilla Conformer}: Vanilla conformer model trained on supervised speech text data only, no text injection.
    
    \textbf{mSLAM}: mSLAM \cite{bapna2022mslam} is a joint speech and text multilingual pretrained model. It consists of a text encoder which is a simple token embedding layer with sinusoidal positional embeddings and layer norm. Text and speech embeddings are concatenated and passed through a multimodal encoder. Text and speech embeddings are aligned through a speech-text matching loss \cite{bapna2021slam}. We borrow mSLAM results as reported in \cite{conneau2022fleurs} where a bigger model is used compared to ASTRA i.e. 600M vs 300M. Their pre-training stage uses 429k hours of wav2vec-BERT \cite{chung2021w2vbert} pretraining, while for ASTRA we do BEST-RQ pretraining on 3M hours of unsupervised speech. This baseline also uses CTC loss instead of RNNT used for ASTRA.
    
    \textbf{Text injection + duration model}: Maestro \cite{chen2022maestro} is a text injection model that relies on learning a common representation for speech and text embeddings by directly minimizing the mean squared error between speech and text embeddings. It also learns a duration model which is used to upsample text embeddings to a comparable length to speech embeddings.
    
    \textbf{Text injection + duration model + VAE}: The above baseline is made stronger by adding a token level VAE to the text encoder which can help augment text embeddings with latent factors such as prosody. This idea is borrowed from \cite{elias2020parallel} where it is applied towards the TTS task. Here we show that a token level VAE when coupled with duration modeling can also help ASR. We train this model, the text-injection baseline and the non text-injection baseline.
    
    All baselines except for mSLAM are initialized from the pretrained mdoel checkpoint described in the pretraining subsection.

%% file: inputfiles/results.tex
\renewcommand{\arraystretch}{2.0} 
\begin{table}[!htp]\centering
    \caption{CER Comparison of various text injection models on FLEURS.}\label{tab:cer-seq}
    \scriptsize
    \begin{tabular}{p{2.0cm} p{2.0cm} p{1.0cm} p{1.0cm}}\hline
    \textbf{Model} &\textbf{Pretraining data} &\textbf{\# params} &\textbf{average CER} \\\hline \hline
    Vanilla Conformer &YT-56-U &300M &13.04 \\ \hline
    w2v-bert-51 \cite{conneau2022fleurs} &VoxPopuli, MLS, CommonVoice, BABEL &600M &14.1 \\ \hline
    mSLAM \cite{conneau2022fleurs} &VoxPopuli, MLS, CommonVoice, BABEL &600M &14.6 \\ \hline
    Text injection + duration model \cite{chen2022maestro} &YT-56-U &300M &13.27 \\ \hline
    Text injection + duration model + VAE &YT-56-U &300M &12.38 \\ \hline
    ASTRA &YT-56-U &300M &12.38 \\ \hline
    \end{tabular}
    \vspace{2.5mm}
\end{table}

On the FLEURS dataset, we are unable to see any CER reduction with the Text injection + duration model compared to the non text-injection baseline. With the ASTRA model presented in this paper, we are able to see a 5\% relative reduction in CER compared to the Vanilla conformer baseline as can be seen in Table \ref{tab:cer-seq}. ASTRA reaches the same CER as the Text injection + duration + VAE model baseline, but without the need to train a duration model. These results show that the performance of upsampling methods is heavily reliant on the quality/sophistication of the duration model. Moreover, it has been shown that alignments generated by RNNT models are delayed \cite{delayedalignment}. This is exactly where our method offers an advantage and will be better suited for modality matching since it implicitly uses the alignments learned by the model itself. This power renders the model invariant to shifted alignments, which implies that one would not need to learn an upsampling and delay the text sequence to try and match it with the speech sequence. Moreover, our method does not rely on any single alignment, instead marginalizes over all alignments and hence, is less reliant on the quality of a single alignment.

\renewcommand{\arraystretch}{2.0} 
\begin{table}[!htp]\centering
    \caption{Model variants and ablations.}\label{tab_ablation}
    \scriptsize
    \begin{tabular}{p{4.5cm} p{1.5cm}}\hline
    \textbf{Model} &\textbf{avg CER} \\\hline \hline
    ASTRA with MSE pointwise consistency &13.45 \\ \hline
    ASTRA with MAE pointwise consistency between shared encoder output of text and speech branches &13.25 \\ \hline
    ASTRA with MAE pointwise consistency between text and speech encoder &12.64 \\ \hline
    + spec aug before RNNT loss on text branch &12.38 \\ \hline
    \end{tabular}
    \vspace{2.5mm}
\end{table}

For the pointwise consistency, we experimented with Mean Absolute Error(MAE) and Mean Squared Error(MSE) and settled on MAE loss due to its improved performance as can be seen in Table \ref{tab_ablation}. We also get a performance boost by adding a SpecAugment layer \cite{Park2019} before the RNNT loss on the text branch. We hypothesize that the SpecAugment layer makes the text RNNT loss more difficult to minimize and helps prevents overfitting on the text corpus. In terms of positioning of the consistency loss, we found it better to place the loss layer at the speech and text encoder output rather than after the shared encoder. We also explored the use of an additional Masked Language Model (MLM) loss~\cite{Bert2019} after the shared encoder, but that did not impact performance.

%% file: inputfiles/conclusion.tex
We introduce ASTRA, a framework which leverages the inherent alignments learned within CTC/RNNT models to learn a joint space between speech and text embeddings and bridge the modality gap for speech-text multi-modal models. We can thus benefit from pure text data on an ASR task, without the need to upsample text embeddings. On the Fleurs ASR task, we show ASTRA has better performance than previous text injection methods, and is on par with a stronger text injection baseline which includes duration model and VAE at the token level. The proposed novel formulation of consistency between modalities as a weighted RNN-T loss allows for easy use in applications requiring LM integration and contextual biasing.